# Epistemological Debt: A Conceptual Framework for AI-Dependent Software Engineering

Frank Ginac, TalentGuard, Austin, TX, USA

Purdue University, West Lafayette, IN, USA

Georgia Institute of Technology, Atlanta, GA, USA

***Abstract*—This paper proposes a conceptual framework for Epistemological Debt, a liability that AI-dependent software engineering creates and conventional engineering metrics miss: the share of a system's behavior that no maintainer can currently explain. The debt accrues in two ways concurrently inside a single workforce. Maintainers with no theory of a system ship generated changes without forming one, and maintainers with a theory stop deriving changes themselves, so the system drifts away from a theory that is still intact. Neither is the attentional lapse the automation complacency literature describes, and the paper explains how to tell them apart. Against technical debt, tacit knowledge, and Naur's theory-building account, the contribution is this: both mechanisms share a cause, and running concurrently, they close the transfer window that earlier accounts of deskilling under automation relied on. The framework is stated as six propositions and two open hypotheses, with an evaluation design that separates the mechanism from the alternatives. Amazon's 2025 and 2026 service incidents illustrate the governance gap the framework predicts; that reading is labeled as interpretation. Proposed interventions are sorted into those that build comprehension and those that only limit risk; none is claimed as validated.**

■ **METRICS OF VELOCITY** and volume dominate the current discourse surrounding Generative AI in software engineering. Leaders frequently cite dramatic reductions in time-to-market and the automation of boilerplate tasks as evidence of a new golden age of productivity.

This focus on functional output obscures a cost that accrues at two levels. For the individual engineer, the derivation that once built a working theory of the system is handed to the model that writes the code, and comprehension decouples from production. For the workforce, that handoff lands on two groups at the same time: maintainers with no theory of the system never build one, and maintainers with one stop doing the work that would keep it current.

Neither loss shows up in conventional engineering metrics. This paper names the resulting liability Epistemological Debt, traces its acceleration under the prompt-driven practice Cito and Bork [1] call Vibe Coding, and states the construct as a short list of propositions, i.e., claims a governance function could act on, and a research design could test. It is a conceptual framework, not a validated model, and it separates three kinds of statements: what the cited literature has established, what this paper hypothesizes, and what it proposes. Its boundary is the organization; what generative tooling does to the public code corpus is a different causal question, noted near the end and left for separate work. As large language models (LLMs) and autonomous coding agents become embedded in the software development lifecycle (SDLC), the industry is accumulating a liability it has not yet started to measure.

## The Emergence of Epistemological Debt

Epistemological Debt is a property of the human-system interface. It is the liability an organization carries as the gap widens between what a software system actually does and what its maintainers can currently explain. Like any debt, it has a principal, i.e., the share of system behavior that no maintainer holds a current theory of, and it comes due at the moment of failure, when remediation demands reasoning that no one on hand can perform.

Generated code may be syntactically correct and pass its unit tests. Still, the maintainer's theory of its state space, failure modes, and transitive dependencies was either never built or no longer matches the system after the change. "Theory" here is meant in Naur's [2] sense: the knowledge a programmer holds that lets them explain what a program does, modify it, and diagnose it when it fails, and that the program text alone does not contain. Polanyi's [3] "tacit dimension" explains why neither condition announces itself: the knowledge is internalized through practice and resists articulation, so a maintainer does not notice its absence or its staleness. The organization's capacity to perform root-cause analysis, or to manage the system during a novel failure with a large blast radius, i.e., one whose effects reach well past the component that failed into the services, data, and users that depend on it, degrades along with it, and none of the instruments the organization routinely consults will register the change.

## Vibe Coding and the Erosion of Abstraction

This debt accelerates with a paradigm shift toward what Cito and Bork [1] call "Vibe Coding." In this model, developers generate complex functional artifacts through natural-language prompts, effectively bypassing the intermediate layers of abstraction. This paradigm collapses the boundary between rapid prototyping and production-grade engineering. Developers become "operators" of opaque systems, accepting probabilistic outputs based on surface-level "vibes" rather than derivations they performed themselves.

## Roadmap of the Analysis

This paper provides an analysis of this liability through four primary lenses:

- *The mechanism*: A taxonomy that separates non-construction, divergence, and inattention, defines the affected groups by what they know of the system rather than by seniority, and sets Epistemological Debt against the technical debt, tacit knowledge, theory-building, and automation complacency literatures.
- *The propositions and evaluation design*: Six propositions, two open hypotheses, an operationalization of the central prediction, the controls it needs, and the results that would refute it.
- *The Amazon case study*: A centrally recommended agentic-tooling policy, the service incidents that followed it, and the company's disputed account of their causes, read as an illustration of the governance gap the framework predicts rather than as evidence of its mechanism.
- *The mitigation framework*: Proposed interventions sorted by what each one is for, i.e., building comprehension, which reduces the debt, or limiting risk, which bounds its cost without reducing it, with an operational definition of the "no-vibe coding" standard.

By naming and scoping the cost of uncomprehended automation, this paper aims to give engineering leaders a liability they can detect and govern. Where it reports an established finding, advances a hypothesis, or proposes an intervention, it says which.

## THE ILLUSION OF UNDERSTANDING

Software engineering is fundamentally an epistemic activity: constructing mental models that map abstract logic to executable reality. Traditionally, the act of writing code, grappling with syntax, logic, and edge cases, serves as the primary mechanism for an engineer to internalize the system's state space. This process builds what Polanyi [3] termed "tacit knowledge": a deep, intuitive understanding of how a system behaves that cannot be easily articulated yet is essential for architectural design and debugging.

### Epistemological Debt

The growing gap between doing and knowing manifests as debt, which this paper calls Epistemological Debt. Technical debt lives in the artifact. Epistemological Debt lives between the code as written and the maintainer's theory of it. Engineers can overestimate their grasp of AI-generated logic because the output is syntactically correct and superficially functional. Becker et al. [4] found that experienced developers working in their own repositories believed AI assistance had sped them up when it had slowed them down; that is a finding about calibration, not system knowledge, cited here for one reason: the affected engineer does not reliably see the gap. The AI masks the underlying complexity, presenting a "black box" solution that works until it does not. When failures occur in edge cases or high-load scenarios, maintainers lack the theory needed to reason about the root cause. They have the "what" (the code) but not the "why" (the derivation).

Modern automation often bypasses a system's essential first-principles logic, leaving developers to generate and deploy complex systems based on probabilistic "vibes" rather than logical, derived proofs. **Figure 1** sketches the accumulation this framework proposes. It is a hypothetical schematic, not a measured trajectory. In the sketch, comprehension tracks complexity through requirements and architecture, where the engineer is still deriving, and the curves part at code generation, where derivation is offloaded. From there, complexity accelerates while comprehension flattens, opening the shaded epistemological debt (ED) region. The evaluation design below is meant to answer whether real trajectories look like this.

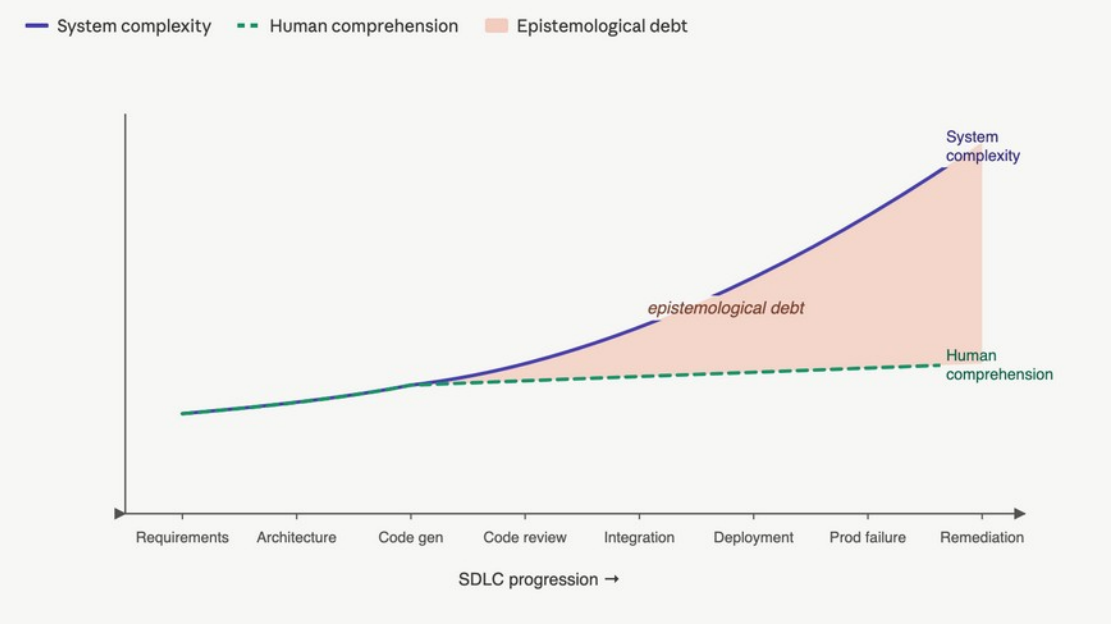

**Figure 1. Hypothetical schematic of Epistemological Debt accumulation across the SDLC. The curves are illustrative, not measured.**

### Taxonomy of Debt

Epistemological Debt does not accumulate uniformly across an engineering organization. Two mechanisms operate concurrently, in different groups, with different origins, onsets, and remedies. However, they produce the same outcome: a maintainer accepts a generated change into production without holding a current theory of the behavior it introduces. The groups are defined by what a maintainer knows of the system, not by seniority. Either a maintainer holds a theory of the specific system, or they do not. A twenty-year veteran who joined the codebase last month is in the second group.

Among maintainers who hold no theory of the system, the mechanism is non-construction. Derivation is handed off before any theory is built, so none gets built. The deficit is one of acquisition: nothing has been lost, because the theory was never there. Among maintainers who hold a theory, the mechanism is divergence. Think of the theory as a map. When its holder stops deriving changes and generated changes land anyway, the map stays exactly as it was while the furniture moves. No cognitive loss is asserted (the map is intact); the mismatch is between a stable theory and a system that has moved on without it. The debt from divergence sits in the code that changed without its holder's derivation, and the change history shows where that is. The two mechanisms are not stages of one process. They run concurrently, on the same codebase, in the same delivery cycle, driven by the same tool. **Table 1** summarizes the distinction.

**Table 1. Two mechanisms of Epistemological Debt accumulation. Evidence and response entries are proposed, not empirically validated.**

| Dimension | Non-construction | Divergence |
|---|---|---|
| **Who is exposed** | Maintainers holding no theory of the system, whatever their tenure | Maintainers holding a theory of a system that changed without them |
| **What is missing** | A theory never formed (acquisition deficit) | Currency: the theory is intact but describes a prior system |
| **Onset** | Immediate, from the first delegated change | Gradual, with each change the holder did not derive |
| **Signal in conventional metrics** | None; delivery velocity typically rises | None; tenure, seniority, and review authority unchanged |
| **Distinguishing evidence** | Comprehension assessed independently of task completion; transfer to an unfamiliar subsystem (task form: Shen and Tamkin [5]) | Same maintainer, unaided: diagnoses failures in code they derived, but not in code that changed without them (identified from the change history); telling them to verify does not close the gap, which rules out inattention. |
| **Comprehension -building response (proposed)** | Construction before delegation: unaided derivation preceding generative tooling | Re-derivation of consequential changes above a blast-radius threshold |

A third pattern is deliberately left out. A maintainer whose theory is present and current may still fail to apply it under load. That is inattention, i.e., the complacency and automation bias that Parasuraman and Manzey [6] describe, and it is not Epistemological Debt. The two are easy to separate in principle. Cue the maintainer to verify. If performance recovers, the lapse was attentional. If it does not, no current theory applies. The distinction matters for governance because senior review, the control most organizations reach for first, only closes the epistemic gap when the reviewer's own theory is current. The automation literature shows that neither training nor instruction to verify eliminates attentional failure in experienced operators. Oversight that rests on senior judgment alone leaves both in place, which is why the safeguards proposed later sit alongside human review instead of replacing it.

Structural artifacts degrade under generative development as well: API contracts drift, and documentation coverage falls. That degradation is real, but it is ordinary technical debt, visible to existing tooling and well served by existing vocabulary. It is not what Epistemological Debt names.

## Two Failures

Consider an engineer who spends hours tracing a race condition through a multi-threaded system. The struggle builds a theory of control flow, memory state, and how the components interact, and when the bug turns up, the engineer knows not just the fix but why it is right. Now consider an engineer new to the system who feeds the stack trace to an LLM instead. Problem-solving shifts from logical deduction to probabilistic induction. The model returns a plausible patch, and the engineer accepts it without deriving it. When the patch fails, the new error state goes back to the model, which announces that it now sees the problem and returns another fix, and they go around, sometimes dozens of times, until the model and the human discover a breakthrough or reach an impasse. Each error state is as opaque as the first, because no theory of the failing behavior was ever built. That is non-construction.

Divergence fails differently. Consider the engineer who built a payment service's retry logic three years ago and still knows it better than anyone. Over the past six months, generated changes reworked the retry path. She reviewed and approved each one, and every test passed. Then customers start seeing duplicate charges. She goes straight to where her theory says duplicates come from, i.e., the idempotency check she designed, and finds it working exactly as intended. One of the generated changes moved where the idempotency key is created, so each retry now arrives with a fresh key, and her check correctly treats every one as a new payment. Her theory still describes the system she built, and it sends her, quickly and confidently, to where the fault would have been in that system.

Both scenarios are illustrations, not evidence, and other explanations, such as time pressure, fit any single instance. The first engineer has no direction to follow; the second follows a precise one that is out of date. Table 1's tests separate them: take the tooling away, compare each maintainer's diagnosis in code they derived against code that changed without them, and cue them to verify.

## Why a New Term

Four strands of prior work have mapped adjacent territory: technical debt, tacit knowledge, theory building, and automation complacency. Each of these literatures locates its phenomenon in one place: a programmer's theory, a practitioner's tacit knowledge, an operator's attention, or, for technical debt, the artifact. Epistemological Debt contributes what none of them supplies.

Technical debt is the nearest neighbor and the most varied. Cunningham's [7] metaphor and its later formalization [8] describe an artifact liability repaid by changing the artifact. The literature since has not treated that liability as uniformly visible or deliberate: Fowler's [9] quadrant separates deliberate debt from inadvertent debt, and static analysis often cannot see architectural debt. So neither visibility nor intent is the dividing line. The dividing line is where the debt lives. Technical debt, in every variant, is a property of the artifact, and changing the artifact retires it at a cost that scales with the code involved. Epistemological Debt is a property of the relationship between maintainer and system. No change to the artifact retires it, and the cheapest artifact change available, regeneration, makes it worse. Paying it down means forming or re-deriving a theory, at a cost that climbs with time and turnover, and it is rarely incurred on purpose, because the maintainer incurring it usually believes their comprehension is intact.

Polanyi [3] supplies the epistemology: knowledge internalized through practice that resists articulation, and whose loss is therefore invisible to the person who has lost it. However, the tacit dimension describes a general property of skilled practice, not a liability that accumulates under a particular production regime.

Naur [2] applied that epistemology specifically to programs. Building on Ryle's [10] distinction between knowing how and knowing that, he argued four decades ago that a program's essential content is a theory held by its programmers, inexpressible in documentation, and that a program "dies" when the team possessing that theory disperses. However, Naur's mechanism is attrition: a theory that existed and was lost with the people who held it. Both of his remedies, consultation with the originating team or discarding the text and re-deriving the solution afresh, presuppose that a theory exists somewhere to be recovered or rebuilt. Generated code violates that premise. No theory of the generated change was ever formed, and the artifact is not its residue. The distinction matters because Naur's preferred remedy is now cheap to execute and useless: regeneration produces the artifact while skipping the byproduct that made re-derivation valuable.

Bainbridge [11] anticipated the two-cohort structure in a single sentence, noting "some concern that the present generation of automated systems, which are monitored by former manual operators, are riding on their skills, which later generations of operators cannot be expected to have." Parasuraman and Manzey [6] then established complacency and automation bias as attentional phenomena: an operator who possesses the required competence reallocates attention away from verification under task load. Their integrated model explicitly excludes cases in which reliance arises because the operator lacks the knowledge required to verify the system at all. Epistemological Debt occupies precisely that excluded region.

What the existing vocabulary does not capture is the unit of analysis and the simultaneity. Simultaneity also answers the question Bainbridge's sentence raises: if his account already describes skilled operators whose successors lack their skills, what does concurrency add? Bainbridge's two cohorts are sequential. The skilled generation is still on the floor while the next one is trained, so there is a transfer window, and workforce planning can schedule around it. Under generative development, the two mechanisms run at the same time, within one workforce, driven by one tool, and what degrades is the transfer: the holders a newcomer would learn from, by deriving alongside them and having their work reviewed, are not deriving either. Three consequences follow that neither half predicts on its own. Transfer fails while pairing, review, and tenure all look normal. Routing generated changes through holders for review, the obvious fix, does not stop divergence, because reviewing a change is not deriving it, and the extra review load feeds inattention. And Naur's two remedies both fail for a structural reason: the originators you would consult are the ones diverging, and the re-derivation you would perform is the activity being offloaded. The two halves do not have equal empirical support, and the framework says so. Shen and Tamkin [5] measured impaired conceptual understanding, code reading, and debugging among developers acquiring an unfamiliar library with AI assistance, i.e., a direct measure of comprehension and transfer, which supports non-construction. For divergence, Becker et al. [4] show only that experienced developers misjudge the tool's effect on their own speed;

they do not measure divergence itself. Whether the two compound within a single workforce is an open question.

The propositions below are the claims this paper advances. The hypotheses are the questions it leaves open.

**P1 (Construct).** Epistemological Debt is an organizational liability equal to the share of a system's behavior that no maintainer currently holds a theory of. That share is the principal, and it comes due when the system fails.

**P2 (Non-construction).** When maintainers who hold no theory of a system hand derivation to generative tooling, changes enter production with no theory formed. The debt grows by the scope of those changes.

**P3 (Divergence).** When maintainers who hold a theory hand derivation to generative tooling, the system changes without their derivation. Their theory stays intact but describes a prior system. The debt grows by the scope of change no holder derived.

**P4 (Concurrency).** Under generative development, P2 and P3 operate on one codebase in one delivery cycle, because one tool displaces derivation for both groups, and the transfer of theory from one group to the other degrades as a result.

**P5 (Invisibility).** Conventional engineering metrics do not register the debt because it is a property of the maintainer-system relationship, not the artifact (test results, velocity, defect rate) or the roster (tenure, headcount).

**P6 (Realization).** The debt is paid at first contact with a failure that no current theory covers. The signature is a longer time to diagnosis, not a longer time to detection, relative to matched controls, with defect rates unchanged.

**H1.** Applying a held theory less often also degrades the theory itself, beyond divergence.

**H2.** Non-construction and divergence compound within a single workforce rather than merely co-occurring.

## Evaluation Design

P6 is the claim that can fail. A drop in conventional metrics does not refute it, because comprehension can decline alongside code quality or delivery speed for unrelated reasons. A longer diagnosis is not unique to this mechanism either; complexity, poor observability, turnover, and inexperience all stretch it. So the prediction is conditional. Among organizations matched on those factors, those with higher AI reliance should take longer to diagnose failures outside any maintainer's current theory, and measured comprehension should account for the difference. **Table 2** operationalizes the pieces.

**Table 2. Evaluation design for proposition P6. A conceptual sketch; no study has been run.**

| Construct | Operationalization | Notes |
|---|---|---|
| **AI reliance** | Share of merged changes generated or substantially drafted by generative tooling, per maintainer and subsystem, from version-control and tool telemetry | Distinguish generation from assistance; the framework concerns delegated derivation |
| **System-specific comprehension** | Assessment independent of task completion, without tooling: explain a subsystem's behavior, predict the effect of a change, diagnose a seeded fault; transfer to an unfamiliar subsystem | Postmortems cannot supply this; assess prior knowledge and its currency directly (task form: Shen and Tamkin [5]) |
| **Unfamiliar-failure exposure** | Classification of incidents by whether the failing behavior lay within any responding maintainer's assessed comprehension at the time | Requires assessment before the incident, or a change history showing who derived the code involved |
| **Time to detection** | Interval from fault onset to first alert or human recognition | Expected to be insensitive to the debt; monitoring determines it |
| **Time to diagnosis** | Interval from detection to identified root cause, as recorded in the postmortem | The outcome variable; expected to lengthen with the debt for unfamiliar failures only |
| **Controls** | System complexity and observability; team turnover; prior incident experience; incident severity | Each lengthens diagnosis independently of comprehension |
| **Falsification** | No comprehension gap between high- and low-reliance groups; or a gap with no matching difference in diagnosis time on unfamiliar failures; or a difference fully explained by controls | Degraded conventional metrics neither confirm nor refute |

Comprehension is assessed directly, not inferred from acceptance rates, productivity, or artifact quality, because none of those measures what the construct names. The design also separates the mechanism from inattention. A maintainer who fails the assessment with the tooling withheld but passes once told to verify has an attentional deficit, not an epistemic one, and the framework predicts a longer diagnosis only for the latter. H1 and H2 require the same assessment repeated over time.

## DELEGATION BEFORE MASTERY

The argument that "AI will do for coding what calculators did for math" relies on a flawed analogy to calculators' automated computation, freeing mathematicians to focus on derivation and higher-order logic. LLMs, conversely, automate the derivation itself. They do not just execute the logic; they write it.

This distinction is critical for foundational training. The "struggle" of learning to code, the frustration of debugging a race condition or optimizing a search algorithm, is not an inefficiency to be eliminated. It is the practice through which a theory of the system gets built.

Survey research is consistent with this. In a CHI '25 study of knowledge workers by researchers at Microsoft and Carnegie Mellon [12], higher confidence in GenAI tools was associated with less self-reported critical thinking, and respondents said they scrutinized AI output less to lower their cognitive load. The same study found that higher confidence in one's own ability went with more self-reported critical thinking. Both are self-reports of effort from knowledge workers at large, not measures of skill or system knowledge among engineers. The step from them to this framework is a hypothesis, not a finding of the study: where derivation is handed off, the scrutiny that came with it goes too.

For maintainers who hold no theory of the system (students included), handing off derivation before mastering it means no theory ever forms, and the maintainer never develops the first-principles understanding required to critique the model's output. One way that shows up at the organizational level is rubber-stamp review, i.e., approval that passes a change without the scrutiny the review exists to provide. Reviewing code is harder than writing it, because it means reconstructing another agent's reasoning, and a reviewer with no theory of the system has nothing to reconstruct it against. Acceptance alone cannot reveal whether a reviewer waves through plausible-looking generated code out of fatigue, trust, or a lack of theory. That is why this framework treats acceptance rates as a measure of posture, not comprehension.

## AMAZON CASE STUDY

A case in point is Amazon. In August 2024, the company reported that its Q Developer agent had migrated tens of thousands of production applications to Java 17, saving an estimated 4,500 developer-years of labor [13], and that developers shipped 79% of the agent's auto-generated code reviews without any additional changes [14]. The figure is company-reported and unaudited, and it says nothing about what the reviewing engineers understood; an unmodified change may simply have been correct. Shipping without modification was reported as an achievement.

In November 2025, an internal memo viewed by Reuters and signed by senior vice presidents Peter DeSantis and Dave Treadwell designated Kiro, a separate spec-driven agentic development tool, as the company's recommended AI-native development tool and closed off support for third-party alternatives [15]. Q Developer and Kiro are different products, but the posture toward them is continuous: acceptance reported as an achievement in 2024 became the centrally designated default in 2025.

Two incidents followed within four months. In December 2025, an AWS service was interrupted in a single region. On March 5, 2026, Amazon's consumer storefront and app were unavailable for roughly six hours, which the company attributed to a software code deployment [16]. On March 10, Amazon convened its weekly retail technology meeting for a deep dive into the outages, and Treadwell wrote to employees that the availability of the site and related infrastructure had not been good recently [17].

Amazon has twice published corrections disputing press accounts that attributed these incidents to AI. On the December interruption, which the account it was correcting had tied to Kiro, the company said the cause was user error, specifically misconfigured access controls that any developer tool could have produced, and that it subsequently required peer review for production access [18]. On the March incidents, which it states did not involve AWS, it said that only one involved AI tooling, that none involved AI-written code, and that the incident arose from an engineer following inaccurate advice an AI tool had inferred from an outdated internal wiki, after which it corrected the source and updated internal guidance [19].

Those are the documented facts. They show a sequence: an agentic tool designated as the recommended path, alternatives closed off, production access available to it, and process gates added after the incidents rather than before them. The reading this paper offers is that the sequence is the governance gap P5 predicts, i.e., nothing in the organization's conventional instruments would have flagged it in advance. That reading is an interpretation, and one of several the facts allow. An engineer acting on inaccurate AI-sourced advice is consistent with incomplete information, time pressure, trust in a source that sounds authoritative, or a weak process, every bit as much as with a gap in that engineer's theory of the system, and the public record does not say what the engineer knew. No claim is made that generated code caused any failure, and none about any individual's knowledge.

On the company's own account, both remedies Amazon reports adopting, peer review for production access and a corrected wiki, are risk-limiting safeguards of the kind proposed below. They are not comprehension-building interventions. Each repairs the environment around the engineer. Neither establishes, nor even tests, whether anyone on hand held a current theory of the system that failed, which is the question this framework says an organization in Amazon's position should be asking. The disagreement between the company and the reporting is over whether AI wrote the code. This framework asks whether humans understood it, and the public record says nothing about it.

## MITIGATION FRAMEWORK

To arrest the accumulation of Epistemological Debt and bound its cost, this paper proposes a mitigation framework with two pillars. Comprehension-building interventions reduce the debt by ensuring a theory is formed or kept current. Risk-limiting safeguards bound what the debt can cost without reducing it. An organization needs both, and neither is validated. The two are easy to confuse: adding a reviewer reduces exposure, but it does not by itself reduce the debt.

*Comprehension-Building Interventions.* These act on P2 and P3 directly. Their defining feature is a bounded human-only assessment: for a defined scope of work, the maintainer derives and defends a change without generative assistance before being allowed to use generative tooling on that scope. That is the operational content of the "no-vibe coding" standard. It does not say that all AI assistance gets in the way of learning (completion, explanation, and search may well help). It says that deriving consequential changes is where a theory forms, and generation substitutes for derivation in a way assistance does not. The standard applies in two settings:

- *Formal academic programs*: Core curricula hold the standard for foundational work. Students cannot critique an LLM's probabilistic outputs against a theory they have not built, so they derive control flow, state-machine dynamics, and transitive dependencies by hand before introducing generative tooling, and assessments verify the derivation rather than the final artifact's correctness.
- *Corporate learning and development*: The standard applies per system rather than per career stage, since the groups are defined by what a maintainer knows of the system. A maintainer new to a codebase, whatever their tenure, derives before generating within it. For maintainers who hold a theory, changes above a blast-radius threshold, i.e., a scope of effect the organization fixes in advance, such as any change that touches a shared service or production data, are re-derived and not merely reviewed, which is the proposed response to divergence in Table 1. Both carry a real cost in delivery time, and the framework does not pretend otherwise.

*Risk-Limiting Safeguards.* These do not build or restore comprehension. They bound what an uncomprehended change can do. They are the right response when the debt already exists and cannot be paid down quickly, and they belong alongside human review rather than behind it, for the reason given in the discussion of inattention: review by a holder whose theory is current is a control; review by anyone else is, at best, a safeguard. Treat generated code with the skepticism due an untrusted, binary-only, third-party component:

- *Specification-to-code traceability*: A written justification that traces the generated logic back to the system's explicit requirements and specifications. The justification limits the reach of an uncomprehended change; whether writing it also builds comprehension is a separate question the framework does not assume an answer to.
- *Invariant checking via property-based testing*: An invariant is a property the system must hold for every input, e.g., a sort routine never loses or duplicates an element, and an account balance never goes negative without an overdraft flag. Frameworks such as Hypothesis and QuickCheck generate many inputs, run the code against them, and report any input that breaks the stated property. That is a test, not a proof. A failure shows the invariant is violated; a pass shows only that no violation turned up among the inputs tried. The safeguard is that a generated change which breaks a stated invariant is caught before deployment, which ordinary example-based unit tests, written for the cases the author thought of, routinely miss.

- *Transitive dependency impact auditing*: A change that adds or upgrades a dependency also pulls in that dependency's own dependencies, several levels deep, and a generated change makes such edits freely because nothing in the prompt said not to. Any AI-assisted refactoring that modifies dependencies therefore triggers an audit of the full tree, not just the packages named in the change, for supply-chain exposure, i.e., packages with known vulnerabilities, packages no longer maintained, and packages whose license or provenance the organization has not approved. The new code compiling and its tests passing says nothing about any of this.

- *Scenario review for consequential changes*: Before deploying a change above a blast-radius threshold, the responsible maintainer states what the change would do under a named failure scenario. The exercise bounds the blast radius and doubles as a comprehension assessment of the kind Table 2 describes, the one place where the two pillars meet.

Labeling each control by its purpose lets an organization see which measures reduce debt and which only contain it. Familiar controls such as review gates, access restrictions, and traceability fall in the second group.

### Beyond the Organization

Human-derived code is the byproduct of the derivation that generative tooling displaces. As the human-authored share of new code falls, the corpus on which future models train shifts toward model-generated code. Shumailov et al. [20] show that indiscriminate recursive training degrades generative models. Gerstgrasser et al. [21] show the degradation stays bounded when synthetic data accumulates alongside real data instead of replacing it. Which regime code corpora follow, and whether the human-authored share can be identified and preserved once provenance is no longer recoverable from the artifact, are open questions this paper does not take up. They fall to the organizations that train models, not to the teams that use a coding assistant. On that trajectory, the organizational liability described here would acquire an ecosystem-scale counterpart. That is a conjecture, left for separate work.

### CONCLUSION

The transition toward AI-dependent software engineering represents the most significant shift in the craft's history since the move from assembly to high-level languages. Earlier abstractions clarified the logic. The current trajectory of generative automation displaces the derivation through which a system's theory is formed. The liability that results, Epistemological Debt, is the share of a system's behavior that no maintainer currently has a theory of. It accrues in two ways at once, within the same workforce: non-construction among maintainers who never build a theory, and divergence among maintainers whose theory the system has moved past. Neither registers on the instruments an organization routinely consults.

This paper has offered a conceptual framework. It has stated the construct as six propositions and two open hypotheses and set it against the technical debt, tacit knowledge, theory-building, and automation complacency literatures. It has also answered what concurrency adds to Bainbridge's account. When newcomers and holders stop deriving at the same time, the holders a newcomer would learn from are not deriving either, and the transfer window that earlier accounts of deskilling under automation relied on closes. The paper has sketched an evaluation design in which comprehension is assessed rather than inferred, and it has separated the proposed mechanism from inattention and the everyday causes of a slow diagnosis. It has read the Amazon incidents as an illustration of the governance gap the framework predicts, and it marks that reading as one interpretation among several.

The mitigation framework proposed here is not a call to abandon automation. It separates the interventions that build comprehension, which reduce the debt, from the safeguards that limit risk, which bound its cost, and it asks engineering leaders to know which is which.

The future of software engineering depends not on how fast we can generate code, but on whether anyone on hand can still explain it. When no one can, the organization holds a liability it can detect and govern.

## REFERENCES

1. J. Cito and D. Bork, “Lost in code generation: Reimagining the role of software models in AI-driven software engineering,” arXiv:2511.02475, 2025.
2. P. Naur, “Programming as theory building,” *Microprocessing and Microprogramming*, vol. 15, no. 5, pp. 253–261, 1985.

3. M. Polanyi, *The Tacit Dimension*. Chicago, IL, USA: Univ. of Chicago Press, 1966.
4. J. Becker, N. Rush, E. Barnes, and D. Rein, “Measuring the impact of early-2025 AI on experienced open-source developer productivity,” METR, 2025.
5. J. H. Shen and A. Tamkin, “How AI impacts skill formation,” arXiv:2601.20245, 2026.
6. R. Parasuraman and D. H. Manzey, “Complacency and bias in human use of automation: An attentional integration,” *Human Factors*, vol. 52, no. 3, pp. 381–410, 2010.
7. W. Cunningham, “The WyCash portfolio management system,” in *Addendum to the Proc. Object-Oriented Programming Systems, Languages, and Applications (OOPSLA ’92 Addendum)*, 1992, pp. 29–30, doi: 10.1145/157710.157715.
8. P. Kruchten, R. L. Nord, and I. Ozkaya, “Technical debt: From metaphor to theory and practice,” *IEEE Software*, vol. 29, no. 6, pp. 18–21, 2012.
9. M. Fowler, “TechnicalDebtQuadrant,” martinfowler.com, Oct. 14, 2009. [Online]. Available: https://martinfowler.com/bliki/TechnicalDebtQuadrant.html
10. G. Ryle, *The Concept of Mind*. London, U.K.: Hutchinson’s University Library, 1949.
11. L. Bainbridge, “Ironies of automation,” *Automatica*, vol. 19, no. 6, pp. 775–779, 1983.
12. H.-P. Lee, A. Sarkar, L. Tankelevitch, et al., “The impact of generative AI on critical thinking: Self-reported reductions in cognitive effort and confidence effects from a survey of knowledge workers,” in *Proc. 2025 CHI Conf. Human Factors in Computing Systems (CHI ’25)*, 2025, Art. no. 1121.
13. Amazon Web Services, “Amazon Q Developer just reached a $260 million dollar milestone,” *AWS DevOps & Developer Productivity Blog*, Aug. 1, 2024. [Online]. Available: https://aws.amazon.com/blogs/devops/amazon-q-developer-just-reached-a-260-million-dollar-milestone/
14. A. Jassy, “One of the most tedious (but critical tasks) for software development teams is updating foundational software,” X, Aug. 22, 2024. [Online]. Available: https://x.com/ajassy/status/1826608791741493281
15. G. Bensinger, “Amazon pushes in-house AI coding tool Kiro over competitors’, memo shows,” Reuters, Nov. 25, 2025. [Online]. Available: https://www.reuters.com/business/retail-consumer/amazon-pushes-in-house-ai-coding-tool-kiro-over-competitors-memo-shows-2025-11-25/
16. A. Rooprai, “Amazon says it fixed issue that led to website outage for thousands,” Reuters, Mar. 5, 2026. [Online]. Available: https://www.reuters.com/business/retail-consumer/amazon-down-thousands-users-us-downdetector-shows-2026-03-05/
17. A. Palmer, “Amazon convenes ‘deep dive’ internal meeting to address outages,” CNBC, Mar. 10, 2026. [Online]. Available: https://www.cnbc.com/2026/03/10/amazon-plans-deep-dive-internal-meeting-address-ai-related-outages.html
18. Amazon, “Correcting the Financial Times report about AWS, Kiro, and AI,” About Amazon, Feb. 20, 2026. [Online]. Available: https://www.aboutamazon.com/news/aws/aws-service-outage-ai-bot-kiro
19. Amazon, “Correcting the Financial Times report about recent Amazon.com service incidents and AI,” About Amazon, Mar. 11, 2026. [Online]. Available: https://www.aboutamazon.com/news/company-news/amazon-outage-ai-financial-times-correction
20. I. Shumailov, Z. Shumaylov, Y. Zhao, N. Papernot, R. Anderson, and Y. Gal, “AI models collapse when trained on recursively generated data,” *Nature*, vol. 631, pp. 755–759, 2024, doi: 10.1038/s41586-024-07566-y.
21. M. Gerstgrasser, R. Schaeffer, A. Dey, et al., “Is model collapse inevitable? Breaking the curse of recursion by accumulating real and synthetic data,” arXiv:2404.01413, 2024.

**Frank Ginac** is the Chief AI Officer at TalentGuard, a human capital management software company in Austin, Texas, where he leads the integration of generative and applied AI into enterprise talent and skills systems. He is a Doctor of Technology student at Purdue University and a Head Instructional Associate at the Georgia Institute of Technology. His research focuses on the socio-technical consequences of AI-dependent software engineering, including epistemological debt and the governance of AI-assisted engineering. Contact him at frank.ginac@talentguard.com.